\documentclass[aps,pra,twocolumn,showpacs,amsmath,amssymb,preprintnumbers,superscriptaddress,10pt]{revtex4-2}
\usepackage{graphicx}
\usepackage{subfig}
\graphicspath{{figures/}}
\usepackage{siunitx}
\usepackage{hyphenat}
\usepackage{braket} 
\usepackage[bookmarks=false]{hyperref}
\usepackage{color}
\usepackage{verbatim}
\usepackage[capitalise]{cleveref}
\crefname{section}{Sec.}{Secs.}
\Crefname{section}{Section}{Sections}
\usepackage{caption}
\usepackage{graphicx}
\usepackage{float} 
\usepackage{CJKutf8}
\usepackage{ragged2e}
\usepackage{diagbox}
\usepackage{booktabs}
\usepackage{array}
\usepackage{tabularx}
\definecolor{pink}{RGB}{255,0,255}

\definecolor{red}{rgb}{0,0,1}

\begin{document}
\newcolumntype{P}[1]{>{\centering \arraybackslash}p{#1}}
\newcolumntype{L}{X}
\newcolumntype{C}{>{\centering \arraybackslash}X}
\newcolumntype{R}{>{\raggedright \arraybackslash}X}

\title{Characterization of Intensity Correlation via Single-photon Detection in Quantum Key Distribution}

\author{Tianyi Xing}
\affiliation{Institute for Quantum Information \& State Key Laboratory of High Performance Computing, College of Computer Science and Technology, National University of Defense Technology, Changsha, 410073, China}

\author{Junxuan Liu}
\affiliation{Institute for Quantum Information \& State Key Laboratory of High Performance Computing, College of Computer Science and Technology, National University of Defense Technology, Changsha, 410073, China}

\author{Likang Zhang}
\affiliation{Hefei National Research Center for Physical Sciences at the Microscale and School of Physical Sciences, University of Science and Technology of China, Hefei, 230026, China}
\affiliation{Shanghai Research Center for Quantum Science and CAS Center for Excellence in Quantum Information and Quantum Physics, University of Science and Technology of China, Shanghai, 201315, China}
\affiliation{Hefei National Laboratory, University of Science and Technology of China, Hefei, 230088, China}

\author{Min-Yan Wang}
\affiliation{Hefei National Research Center for Physical Sciences at the Microscale and School of Physical Sciences, University of Science and Technology of China, Hefei, 230026, China}
\affiliation{Shanghai Research Center for Quantum Science and CAS Center for Excellence in Quantum Information and Quantum Physics, University of Science and Technology of China, Shanghai, 201315, China}
\affiliation{Hefei National Laboratory, University of Science and Technology of China, Hefei, 230088, China}

\author{Yu-Huai Li}
\affiliation{Hefei National Research Center for Physical Sciences at the Microscale and School of Physical Sciences, University of Science and Technology of China, Hefei, 230026, China}
\affiliation{Shanghai Research Center for Quantum Science and CAS Center for Excellence in Quantum Information and Quantum Physics, University of Science and Technology of China, Shanghai, 201315, China}
\affiliation{Hefei National Laboratory, University of Science and Technology of China, Hefei, 230088, China}

\author{Ruiyin Liu}
\affiliation{Institute for Quantum Information \& State Key Laboratory of High Performance Computing, College of Computer Science and Technology, National University of Defense Technology, Changsha, 410073, China}

\author{Qingquan Peng}
\affiliation{Institute for Quantum Information \& State Key Laboratory of High Performance Computing, College of Computer Science and Technology, National University of Defense Technology, Changsha, 410073, China}

\author{Dongyang Wang}
\affiliation{Institute for Quantum Information \& State Key Laboratory of High Performance Computing, College of Computer Science and Technology, National University of Defense Technology, Changsha, 410073, China}

\author{Yaxuan Wang}
\affiliation{Institute for Quantum Information \& State Key Laboratory of High Performance Computing, College of Computer Science and Technology, National University of Defense Technology, Changsha, 410073, China}


\author{Hongwei Liu}
\affiliation{China Information Technology Security Evaluation Center, Beijing, 100085, China}

\author{Wei Li}
\affiliation{Hefei National Research Center for Physical Sciences at the Microscale and School of Physical Sciences, University of Science and Technology of China, Hefei, 230026, China}
\affiliation{Shanghai Research Center for Quantum Science and CAS Center for Excellence in Quantum Information and Quantum Physics, University of Science and Technology of China, Shanghai, 201315, China}

\author{Yuan Cao}
\affiliation{Hefei National Research Center for Physical Sciences at the Microscale and School of Physical Sciences, University of Science and Technology of China, Hefei, 230026, China}
\affiliation{Shanghai Research Center for Quantum Science and CAS Center for Excellence in Quantum Information and Quantum Physics, University of Science and Technology of China, Shanghai, 201315, China}
\affiliation{Hefei National Laboratory, University of Science and Technology of China, Hefei, 230088, China}

\author{Anqi Huang}
\email{angelhuang.hn@gmail.com}
\affiliation{Institute for Quantum Information \& State Key Laboratory of High Performance Computing, College of Computer Science and Technology, National University of Defense Technology, Changsha, 410073, China}

\date{\today}

\begin{abstract}
One of the most significant vulnerabilities in the source unit of quantum key distribution~(QKD) is the correlation between quantum states after modulation, which shall be characterized and evaluated for its practical security performance. In this work, we propose a methodology to characterize the intensity correlation according to the single-photon detection results in the measurement unit without modifying the configuration of the QKD system. In contrast to the previous research that employs extra classical optical detector to measure the correlation, our method can directly analyse the detection data generated during the raw key exchange, enabling to characterize the feature of correlation in real-time system operation. The basic method is applied to a BB84 QKD system and the characterized correlation significantly decreases the secure key rate shown by the security proof. Furthermore, the method is extended and applied to characterize the correlation from the result of Bell-state measurement, which demonstrates its applicability to a running full-scheme MDI QKD system. This study provides an approach for standard certification of a QKD system.
\end{abstract}
\maketitle

\section{Introduction}
\label{sec:intro}
Quantum key distribution~(QKD) is promising for distributing secret keys between two remote parties, which makes information-theoretic security communication possible based on the fundamental principles of quantum mechanics~\cite{1984Quantum, 1991Ekert}. 
However, imperfections in practical devices that disclosed in QKD systems may contribute to security loopholes and threaten its practical security~\cite{lydersen2010hacking,huang2019laser,PhysRevApplied.13.034017,wu2020hacking, 7571108, e24020260, PhysRevA.106.033713}. The advent of measurement-device-independent~(MDI) QKD has effectively addressed the security vulnerabilities in the measurement unit~\cite{lo2012measurement,yin2016measurement,cao2020long,gu2022experimental,wang2019asymmetric,li2023free}, paving the way for substantial advances in QKD systems. Thus, the loopholes in the measurement unit have been eliminated, the research community begins to focus on the security loopholes in the source unit~\cite{huang2019laser,PhysRevApplied.13.034017,ponosova2022protecting,peng2024security,lovic2023quantified,lu2021intensity,huang2023characterization}. \par

One of the loopholes in the source unit is the intensity correlation among quantum states due to the memory effect of optical modulators and the cross-talk of their electrical driver. That is, the intensity of quantum states prepared in the previous rounds influence that of the current one, especially in high-speed QKD systems~\cite{yoshino2018quantum,grunenfelder2020performance,roberts2018patterning}. This phenomenon violates the basic assumption in most QKD security proofs that all the quantum states emitted by the QKD transmitter are independent and identically distributed~\cite{hwang2003quantum,lo2005decoy,wang2005beating,ma2005practical}. Consequently, the loophole provides an opportunity for an eavesdropper to learn the secret key information according to the correlation, thereby compromising the security of the system~\cite{lo2007security,Pereira2019quantum,pereira2020quantum,zapatero2021security,sixto2022security,pereira2023modified,Lu:23}.\par

To address the threat of correlation in intensity modulation, some theoretical methods are proposed to provide the security analysis structure~\cite{zapatero2021security,sixto2022security,pereira2023modified} and to reduce the impact of the correlation~\cite{PhysRevLett.131.110802,Lu:22}. These methods consider the intensity correlation in the analytical framework in order to bridge the gap between practice and theory. Furthermore, there are some experimental studies to calibrate the correlation in the QKD systems~\cite{yoshino2018quantum,kang2022patterning}.
It is important to note that these experiments usually characterize the correlation of optical pulses by the classical optical detector, assuming the correlation results are consistent with those at the single-photon-level. This method is inconsistent with the original configuration of a QKD system that typically employs single-photon detectors. That is, it is required to modify the system by reducing the optical attenuation in the transmitter and applying an additional classical optical detector to measure and evaluate the intensity correlation. Unfortunately, this modification might not be feasible for the mature and compact QKD system provided by the third party. In addition, security certification of the QKD system shall be applicable to the one in the initial working mode without reducing the attenuation in the transmitter~\cite{ISO/IEC23837-1}.\par

This paper proposes a methodology to characterize the intensity correlation in the QKD system, without modifying its configuration. The intensity correlation can be characterized by analyzing the detection data from the single-photon detectors (SPDs) originally utilized in the measurement unit of the QKD system. To demonstrate the generality and applicability, this methodology is applied to characterize the intensity correlation of the quantum state modulation in a 250 MHz BB84 QKD system with time-phase encoding and a 1.25 GHz MDI QKD system with polarization encoding. The correlation parameters obtained from the BB84 QKD system are employed to the available security proof to illustrate the decreased secret key rate, while the security proof of MDI QKD is not yet fully developed.  \par

This study shows that the generalized method is applicable to the original configuration of QKD systems, without the need to modify the QKD system to reduce attenuation or to add any extra measurement equipment. It is suitable for different systems employing various QKD protocols, as demonstrated by our work on a BB84 QKD system and a MDI QKD system. Moreover, the method is capable of characterizing intensity correlation by directly processing the data obtained during the raw key exchange, which allows to dynamically monitor the correlation in real time and calculate the secure key rate simultaneously.  \par

This paper is organized as follows. In Sec.~\ref{sec:Methodology}, we analyze the relationship between the click probability and the detection efficiency and present the general methodology to characterize the intensity correlation, according to the detection result of the SPDs. Next, in Sec.~\ref{Application on BB84 QKD} and \ref{Application on MDI QKD}, we apply the method to a BB84 QKD system and a MDI QKD system, respectively, which leads to the observation of correlation fluctuations in the systems. Furthermore, according to the correlation results presented in Sec.~\ref{Application on BB84 QKD}, we analyze the effect of the correlation on the security of the BB84 QKD system in Sec.~\ref{Effect on the Security of QKD}, followed by the conclusions in Sec.~\ref{Conclusion}.

\section{General Methodology}
\label{sec:Methodology} 

Our methodology aims to provide security evaluators with a feasible approach to estimate the intensity correlation of the QKD system, certifying its practical security performance. That is, the application scenario of the proposed methodology is that the whole QKD system under test shall be in a fully trusted certification room and no modification on the configuration of the tested QKD system is needed. Since this methodology is proposed to characterize the intensity correlation of quantum states according to the detection data from the SPDs, it is first necessary to analyze the relationship between the intensities of quantum states and the count rates of the SPDs. Assume that the detection efficiency of a single photon in the QKD system under test is $\eta$, including the transmission efficiency of the quantum channel and the detection efficiency of the SPDs, and the mean photon number emitted by the source unit of the QKD system is $m$. The photon number emitted by the weak coherent source follows the Poisson distribution $P(n)= m^n e^{-m}/n!$, in which $n$ represents the photon number. Therefore, the probability $P_{NoClick}$ that the SPD in the QKD system does not click is 

\begin{align}
P_{NoClick}=\sum_{n=0}^NP(n) (1-\eta)^{n},
\end{align}

i.e.~all the photons are not detected after transmission. Thus, the probability $P_{Click}$ that the SPD clicks is

\begin{align}
P_{Click} &=\sum_{n=0}^NP(n)(1-(1-\eta)^{n})
\\&=1-\sum_{n=0}^N \frac {m^n(1-\eta)^n}{n!}\cdot e^{-m}\nonumber
\\&=1-e^{-\eta m}\nonumber
\end{align}

\begin{figure}[h]
\centering
\includegraphics[width=\linewidth]{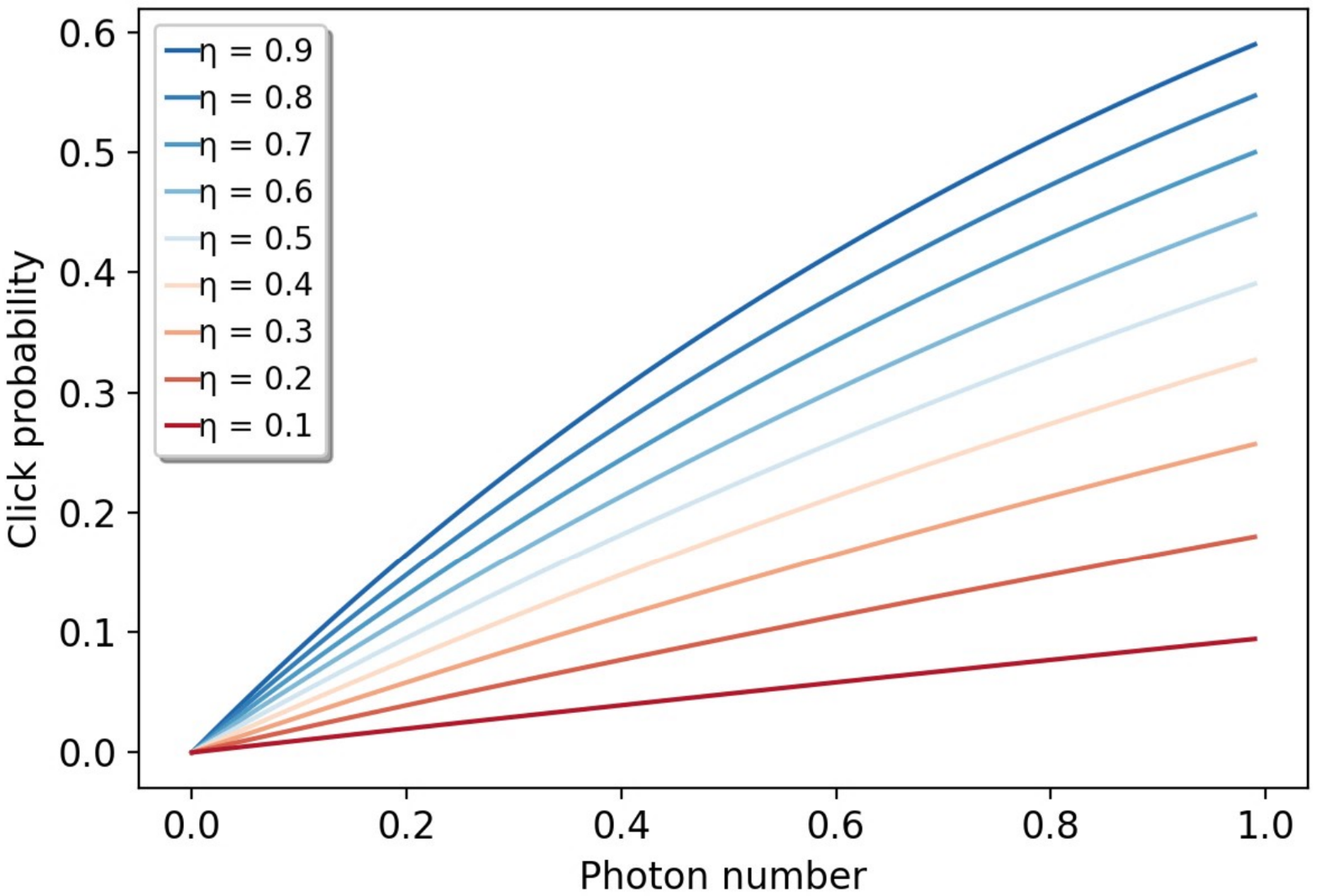}
\captionsetup{font={small,stretch=1.25},justification=raggedright}
\caption{{\bfseries The click probability for a single-photon detector.} \justifying{When the mean photon number $m \in (0,1)$, the click probability is positively related to the detection efficiency~($\eta$).}}
\label{click probability}
\end{figure}

The relationship between the click probability and the detection efficiency $\eta$ is simulated as in Fig.~\ref{click probability}, when the mean photon number~$m$ is less than 1. The source unit of the QKD system attenuates the light to the single-photon scale, so the mean photon number $m$ is controlled to be less than 1. As illustrated, when $m \in (0,1)$, the curves can be considered as a linear relationship between the click probability and the mean photon number. It is indicated that the click probabilities of the SPDs can be used to linearly represent the intensity of single-photon-level states, and thus the intensity correlation of the optical pulse can be explored by the click rate of the SPDs.

Once the linear relationship between SPD's click probability and mean photon number has been verified, the click rate in the measurement unit of a QKD system can be employed to represent the average intensity of the corresponding quantum states, which characterizes the intensity correlation. The outcome of the intensity correlation can be obtained through five stages as follows. \par
\emph{Stage 0 Check and define parameters}\par
\begin{figure}[h]
\centering
\includegraphics[width=\linewidth]{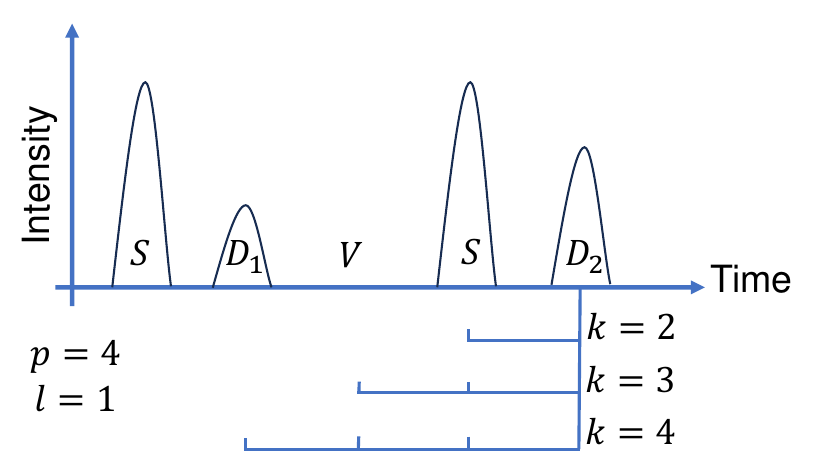}
\captionsetup{font={small,stretch=1.25},justification=raggedright}
\caption{{\bfseries Schematic for the intensity of optical pulses emitted from a single source.} \justifying{The pulses show a QKD system with a single source ($l=1$) and four intensities produced ($p=4$). Different lengths of the correlated intensity group are demonstrated with different values of $k$.}}
\label{Method Fig}
\end{figure}

Before analyzing the correlation among the intensities of quantum states, it is vital to describe the parameters of the intensity correlation in modulation via checking the fundamental settings of the QKD system and defining the correlation length. Two basic parameters of the tested QKD system related to the correlation are the number of distinct intensities of the pulse, denoted as $p$, and the number of source units, denoted as $l$. The correlation
length is denoted as $k$, ranging from 1 to $N$ as an
integer, which defines that the intensities of the $k - 1$ previous pulses influence the intensity of the $k$-th pulse. Figure~\ref{Method Fig} illustrates a series of quantum-state pulses produced by a QKD system containing one source ($l=1$) and four intensities ($p=4$). The system is configured to produce pulses in the signal state ($S$), the decoy states ($D_1$ and $D_2$), and the vacuum state ($V$). The correlation length is considered to be $k$ = 2, 3, or 4, presenting the correlated modulation pattern.\par

\emph{Stage 1 Grouping}\par
After obtaining the parameters of the QKD system under test, the pulses with the same combination of intensity settings need to be grouped in order to analyze the correlation. It is noted that this method is designed to analyze one target source. In the case of multi-source protocols, one of the $l$ sources is selected as the target source to analyze its modulation correlation. Consequently, the analysis of multiple sources can be performed simultaneously by applying the method to each source. For a given set of $p$ intensities, the length of the modulation patterns is considered to be $k$, which results in a total of $p^k$ intensity groups. For the specific intensity of the $k$-th pulse, there are $p^{k-1}$ groups of different intensity combinations from the first to the $(k-1)$-th pulses. Taking $k=2$ in Fig.~\ref{Method Fig} as an example, there are 16 ($4^2$) groups ($a_1a_2$, for $a_1,a_2 \in \{S,D_1,D_2,V\}$). For the signal state $S$ as the $k$-th pulse, there are four corresponding groups of intensity combinations ($VS$, $D_1S$, $D_2S$ and $SS$). 
All the parameters required in Stage 0 and Stage 1 are available, once the protocol and the structure of the QKD system are determined. Thus, these two stages can be accomplished before the QKD system is operational.\par

\emph{Stage 2 Transmission counting}\par
The second stage is to count the number of transmissions for each group in the transmitted pulse sequence, during the key exchange. When there is a correlated modulation pattern, $a_1a_2$, in the transmitted pulse sequence, the transmission count for the corresponding group should be added by one, denoted as $T_i$. $i$ represents a specific group, selected from intensity groups with a total of $p^k$. In the case that all the pulses emitted by the source unit are shown in Fig.~\ref{Method Fig}, $i$ can be $D_1V$, $VD_2$, $D_2S$ and $SD_1$, all of which are included in the 16 groups mentioned in Stage 1. For $T_i$, there is a single count of $D_1V$, $VD_2$, $D_2S$, and two counts of $SD_1$ for this source. \par

\emph{Stage 3 Click collecting}\par
The third stage is to collect the detection clicks of each group, $C_i$. $i$ represents the group that comprises $k$ pulses. Upon detecting the $k$-th pulse, the detection collection ($C_i$) of the group is increased by one, without the need to detect the previous $k-1$ pulses.
Given that all the pulses in Fig.~\ref{Method Fig} are detected, the clicks of the detection should be collected into their own $C_i$, where $i= D_1V$, $VD_2$, $D_2S$ or $SD_1$. The emitted quantum states needed in Stage 2 and the detection clicks needed in Stage 3 can be conducted after the raw key exchange.\par

\emph{Stage 4 Calculation and comparison}\par
The final stage is to calculate and compare the mean click rate for each group, $R_i=C_i/T_i$, which can be calculated immediately after the Stage 3. The $R_i$s of the groups with the same intensity setting at the $k$-th pulse can be compared to illustrate their correlations with different previous intensity settings. The greater the relative fluctuation of the click rates among the groups with the same intensity at the $k$-th pulse is, the higher correlation among quantum states is.\par
It is notable that the intensity correlation is the major but not the only reason for the discrepancy in click rates. The dark count and the after-pulse in the detectors may also may cause the different click rate for the same intensity choice, which shall be a minor factors that have not modelled in details in the methodology. After presenting the methodology of characterizing intensity correlation analysis at the single-photon level, we apply it to BB84 and MDI QKD systems verifying its generality.

\section{Application on BB84 QKD}
\label{Application on BB84 QKD}
\begin{figure*}[!htbp]
\centering
\includegraphics[width=0.8\linewidth]{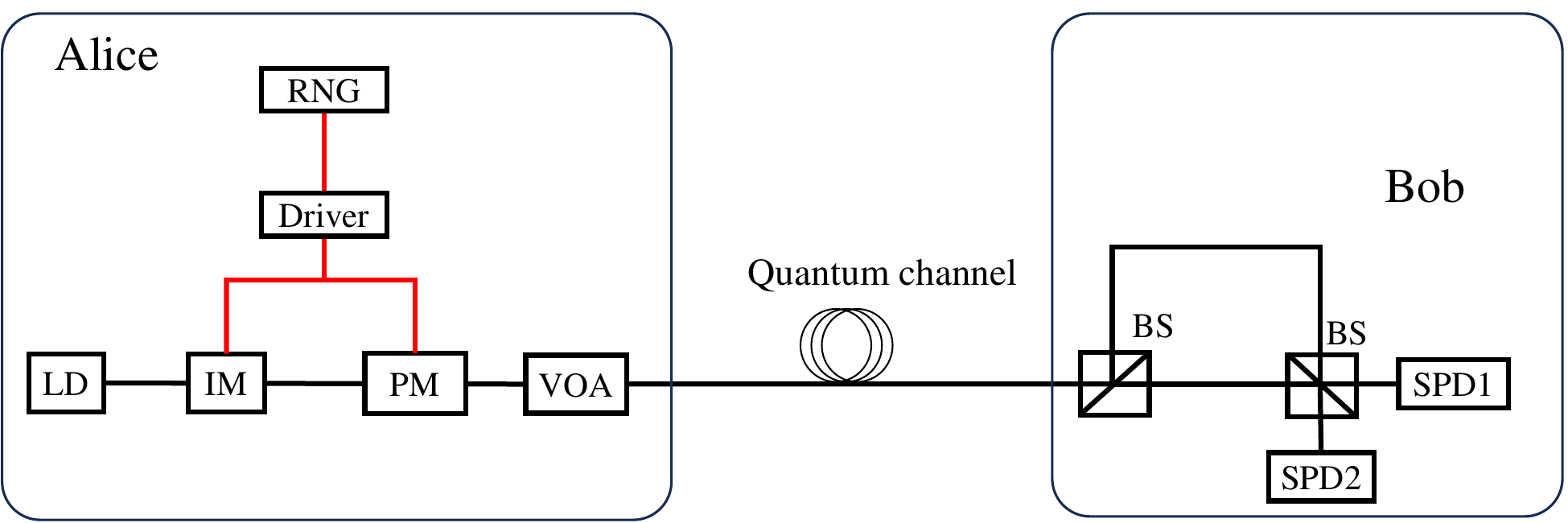}
\captionsetup{font={small,stretch=1.25},justification=raggedright}
\caption{{\bfseries The schematic setup of the time-phase-encoding BB84 QKD system under test.} \justifying{The source unit generates the pulse with the laser~(LD), which is modified by an intensity modulator~(IM), a phase modulator~(PM), and a variable optical attenuator~(VOA). The random number generator~(RNG) produces the random number to the driver, which is used to encode the quantum states by controlling Alice's PM and IM. In the measurement unit, Bob uses an asymmetric Mach–Zehnder interferometer~(MZI) with two beam splitters~(BSs) and two SPDs to measure the quantum states. The electronic connections are shown in red and the optical connections are shown in black.}}
\label{Model BB84}
\end{figure*}

Applying the methodology outlined above, we analyze the intensity correlation in a BB84 system and present the result as follows. The entire time-phase-encoding BB84 system is schematically shown in Fig.~\ref{Model BB84}. Alice produces the quantum states via the laser~(LD), the intensity modulator~(IM), phase modulator~(PM), and the variable optical attenuator~(VOA). The random number generated by the random number generator~(RNG) is used to encode the quantum state by the driver of the phase modulator and the intensity modulator. There are four intensity states applied in the tested BB84 system, with $V$ for the vacuum state, $D_1$ and $D_2$ for two decoy states with different intensities, and $S$ for the signal state. A random number string is used to encode 10000 quantum states, with a ratio of roughly $V:D_1:D_2:S=3:7:35:55$. The quantum-state sequence is transmitted repeatedly by Alice for \num{1.5e7} times. As for detection, Bob uses an asymmetric Mach–Zehnder interferometer~(MZI) to detect the pulses.\par

\begin{figure}[h]
\centering
\includegraphics[width=\linewidth]{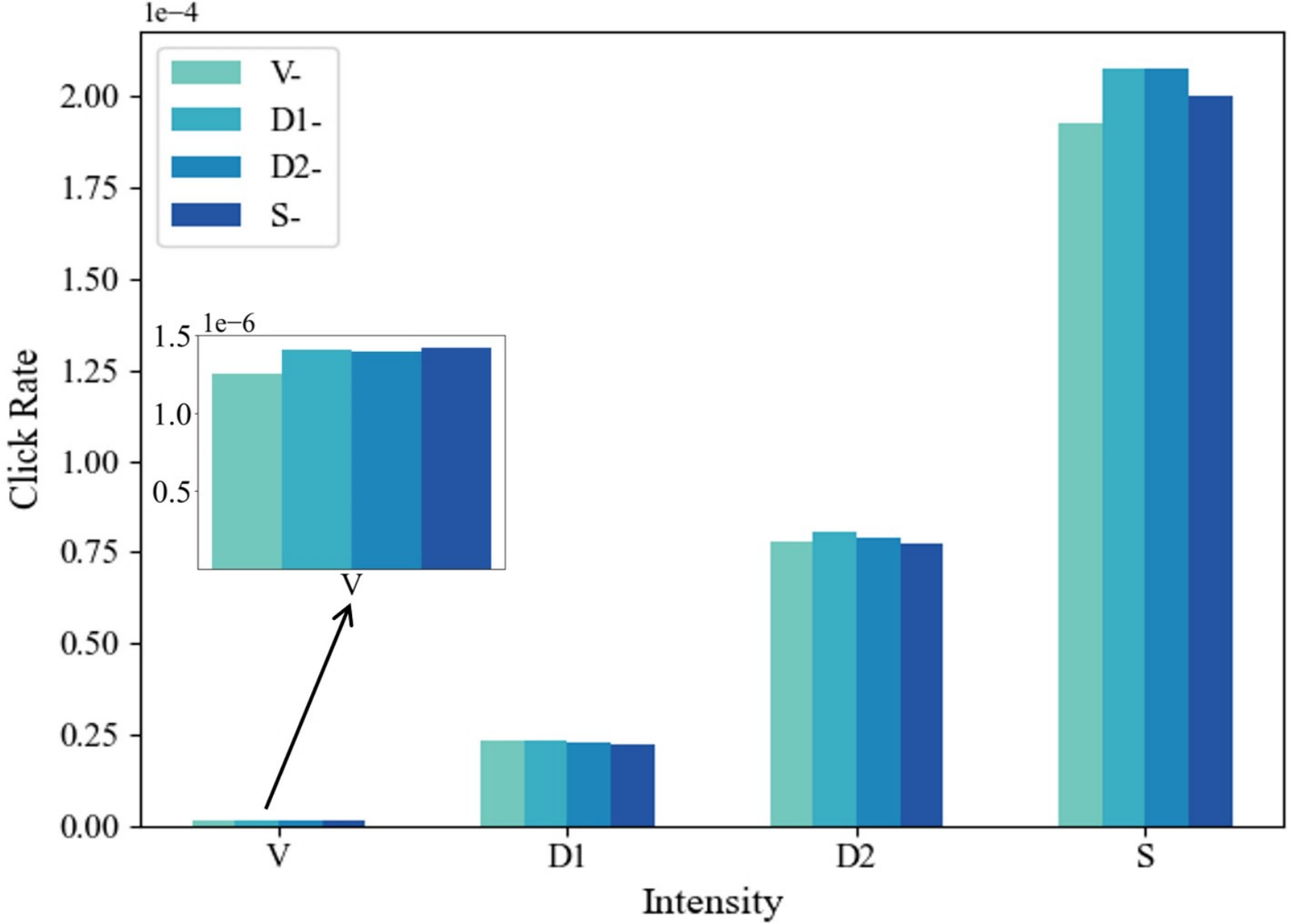}
\captionsetup{font={small,stretch=1.25},justification=raggedright}
\caption{{\bfseries The correlation result for BB84 QKD.} \justifying{There are four intensities, $V$ for the vacuum state, $D_1$, $D_2$ for the decoy state and $S$ for the signal state. For the same intensity, the click rates varies with different previous intensities, which demonstrates the intensity correlation between the former and latter pulses.}}
\label{Result for BB84}
\end{figure}

In accordance with the Stage 0, the BB84 system has one source with $l=1$ and there are four distinct intensities ($a=V, D_1, D_2, S$), $p=4$. In this work, in order to simplify the calculation, we consider the neighbour-nearest correlation with $k=2$ in the BB84 system. Consequently, there are 16 ($4^2$) groups in Stage 1 ($a_1a_2$, for $a_1,a_2=V, D_1, D_2, S$) for this QKD system. Regarding the transmission counting in Stage 2, the total number of transmissions $T_i$ for a group can be achieved by multiplying the number of transmissions $G_i$ for the group in the sequence via the number of repetitions $r$, i.e. $T_i=G_i*r$. In the 10000-long sequence, taking signal state $S$ at the $k$-th pulse as example, the group of $VS$ transmits \num{155} times, the group of $D_1S$ transmits \num{1927} times, the group of $D_2S$ transmits \num{380} times, and the group of $SS$ transmits \num{3033} times in each sequence. This sequence is transmitted \num{1.5e7} times repeatedly during the entire raw key exchange procedure. Consequently, after multiplying the repetition times, $T_{VS}=$\num{2.325e9}, $T_{D_1S}=$\num{2.8905e10}, $T_{D_2S}=$\num{5.7e9}, and $T_{SS}$=\num{4.5495e10} in total~\cite{note1}. Accordingly, in Stage 3, the detection result $C_i$ for $S$ at the $k$-th pulse includes $C_{VS}=$\num{448412}, $C_{D_1S}=$\num{6001994}, $C_{D_2S}=$\num{1183455} and $C_{SS}=$\num{9101922}. Finally, in Stage 4, the click rate $R_i$ = $C_i/T_i$ is calculated for all groups, as shown in Fig.~\ref{Result for BB84}. 

\begin{figure*}[t]
\centering
\includegraphics[width=\linewidth]{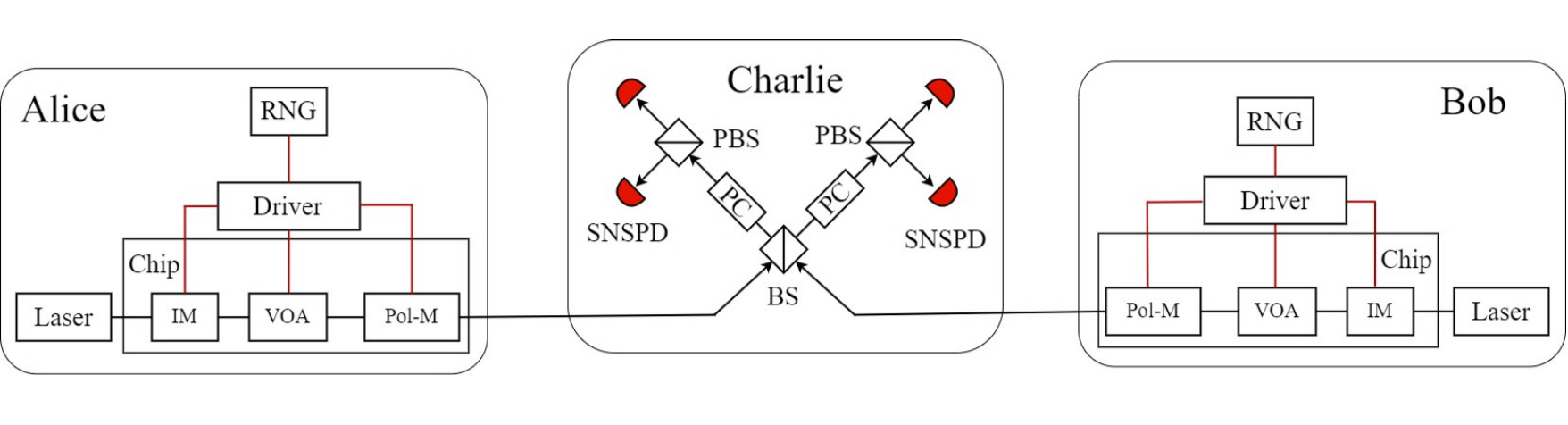}
\captionsetup{font={small,stretch=1.25},justification=raggedright}
\caption{{\bfseries The schematic setup of MDI QKD system.} \justifying{Alice and Bob use the RNGs to generate the random number, used to encode the quantum state via the modulators on the chip. The untrusted third part, Charlie, deploys the Bell-state measurement in the measurement unit with the beam splitter, the polarization beam splitters, and the superconducting nanowire SPDs. The electronic connections are shown in red and the optical connections are shown in black.}}
\label{Model for MDI}
\end{figure*}
In the ideal scenario where the intensities of pulses are independent without correlation, the click rates corresponding to groups with the same intensity at the $k$-th pulse should be the same. However, it is evident that there is a gap between the click rates of the groups with the same intensity at the $k$-th pulse. For $D_1$, $D_2$ and $S$ acting as the $k$-th pulse, their click rates decrease when the previous intensity is $V$ or $S$, while increase when that is $D_1$ and $D_2$. For $V$ at the $k$-th pulse, its click rate increases as the previous intensity increases from $V$ to $S$. Comparing the different $k$-th pulses, the signal state $S$ has the largest absolute discrepancy, approximately \num{1.5e-5} (7\% in relative), while the maximal relative fluctuation is roughly 12\% for the vacuum state $V$. For $D_1$ and $D_2$, the mean click rates are \num{2.3e-5} and \num{7.8e-5} respectively, with both exhibiting a 4\% relative fluctuation. 
The discrepancies indicate that the intensity of the quantum state is influenced by the correlation of the previous one, which introduces a potential security loophole.\par

\section{Application on MDI QKD}
\label{Application on MDI QKD}
The system under consideration in this section is a photonic-chip-based MDI QKD system operating at a repetition frequency of \SI{1.25}{GHz}, as illustrated in Fig.~\ref{Model for MDI}. Alice and Bob are the two source units in the MDI QKD system. The optical pulses generated by the laser are modulated by the integrated chip, which includes an intensity modulator~(IM), a variable optical attenuator~(VOA), and a polarization modulator~(Pol-M), all of which are controlled by the driver. The RNG generates the random number, encoded by the driver and the optical modulators into the quantum states. The Bell-state measurement is performed by the untrusted third party, Charlie, who employs a beam splitter~(BS), two polarization controllers~(PCs), two polarization beam splitters~(PBSs) and four superconducting nanowire single-photon detectors~(SNSPDs). \par

During the raw key exchange, Alice and Bob utilize two distinct 8192-long random number strings to encode the quantum states. In each string, $0$/$1$ correspond to the signal state $s$ in $H/V$ polarization, $2$ and $3$ correspond to the first decoy state $\mu$ in $+/-$ polarization, $4$ and $5$ correspond to the second decoy state $\nu$ in $+/-$ polarization, and $6$ and $7$ correspond to the vacuum state $\omega$ in $+/-$ polarization. Alice and Bob emit the pulses according to their own random number strings respectively and repeat approximately \num{1.6e6} times. The ratio for the intensity states in Alice's/Bob's string is $\omega : \nu : \mu :s=1:1:3:7$. The measurement unit of Charlie records the arrival time of each pulse and the serial number of the clicking detectors.\par

We first select Alice or Bob in a MDI QKD system as a target source. We then analyse the correlation characteristics of the target source with an assumption that the quantum state sent by the other source is stable with no correlation for simplicity. Regarding this target source, the general correlation method described in Sec.\ref{sec:Methodology} indicates that the MDI QKD system has four intensities ($s$, $\mu$, $\nu$, $\omega$) for the decoy-state protocol, $p=4$ in Stage 0. As an application example of the methodology, the neighbor-nearest correlation, $k=2$, is considered in this MDI QKD system for simple demonstration and data processing. Consequently, in the Stage 1, there are 16~($4^2$) groups for each source unit, Alice/Bob, in the MDI QKD system. Without prejudice to the generality, we choose Alice as the target source in our work.\par

As for Stage 2, the number of transmissions for groups will be counted according to the random number string and multiplied by the repeat times. In this MDI QKD system, the signal state only employs Z basis with $H/V$ polarization, while the others only employ X basis with $+/-$ polarization. According to the MDI QKD protocol, Alice and Bob sending the states at the same basis would perform the valid Bell-state measurement. Thus, when the quantum states of the two pulses from Alice and Bob have the same polarization basis, the pulse will be counted in Stage 2. In the experiment, referring to the polarization selections corresponding to the random number, 2802 pulses in total among the string are filtered and the number of transmissions for each group ($G_i$) in the string is shown in Table.~\ref{transfers Table}. \par 
\begin{table}[h!]
  \caption{The number of transmissions for groups, $G_i$, in the random number string of the MDI system.}
  \centering
  \renewcommand{\arraystretch}{1.2}
   \begin{tabularx}{0.5\linewidth}{P{1.45cm}| C C C C}
    \hline
         \diagbox{Pre}{Cur} &  $\omega$ & $\nu$ &  $\mu$ & $S$ \\
        \hline
         \centering $\omega$ & 19 & 59 & 27 & 109 \\
         \centering $\nu$& 75 & 213 & 65&353 \\
         \centering $\mu$& 18 & 54 & 26 & 126 \\
         \centering $S$ & 163& 500& 157 & 838\\
    \hline
    \end{tabularx}
\label{transfers Table}
\end{table}

In Stage 3, since the detection in MDI QKD requires Charlie to perform the Bell-state measurement, the valid clicks ($C_i$) in Stage 3 after two-photon interference are determined by the post-selected result of either the $\ket{\Psi^+}$ or the $\ket{\Psi^-}$~\cite{lo2012measurement}. Thus, the detection results in the measurement unit, of which only the coincidence count conforming to the $\ket{\Psi^+}$ or the $\ket{\Psi^-}$, will be counted in $C_i$. However, in practice, the raw data obtained by Charlie is the coincidence count result between Alice's and Bob's pulses in low average photon level. Consequently, the actual click rate of the coincidence count in MDI QKD system is considerably lower than that of BB84 QKD, which thereby lacks adequate samples of the click and experiences high statistic fluctuations for correlation analysis. To address this issue, we define a detection cycle as the process of detecting the pulses of the random number string once and search the cross-cycle coincidence from the previous cycle to the subsequent cycles. The cross-cycle coincidence counts are collected to increase the number of detection clicks ($C_i$) in Stage 3. \par

\begin{figure}[t]
\centering
\includegraphics[width=0.45\textwidth,height=0.5\textwidth]{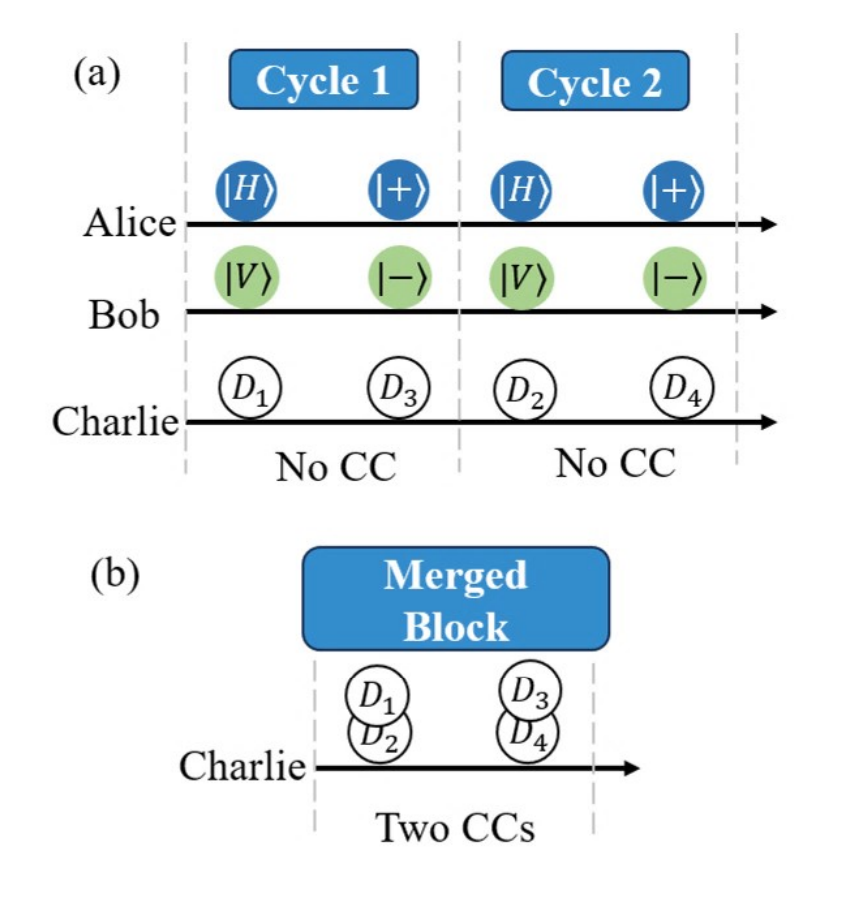}
\captionsetup{font={small,stretch=1.25},justification=raggedright}
\caption{{\bfseries The cross-cycle method.} \justifying{(a) Alice and Bob respectively and repeatedly transmit their own quantum states. Ordinarily, Charlie can not obtain the coincidence count~(CC) in Cycle 1/2 owing to the low detection efficiency of the system. (b) If we consider Cycle 1 and Cycle 2 as a merged block, in the first slot, $D_1$ and $D_2$ conform to the Bell-state measurement, where there is a CC. Similarly, $D_3$ and $D_4$ can also perform a CC, in the second slot. Consequently, there are two CCs in this merged block.}}
\label{Method for cross-cycle}
\end{figure}

The specific method of the cross-cycle coincidence is as follows. Without loss of generality, it is assumed that the transmission channel from the source unit to the detector is lossless, and all the loss occurs in the detection process of receiver, Charlie.  
Thus, the MDI QKD system detects the pulses multiple times, which is considered to be the sampling process of a lossless channel. In order to obtain an accurate detection result in this conditions, it is necessary to collect the result among multiple detection cycles. As illustrated in Fig.~\ref{Method for cross-cycle}(a), in this MDI QKD system, Alice and Bob individually and repeatedly emit their sequences of the randomly-encoded quantum states. It is assumed that Alice transmits $\ket{H}$ in the first slot and $\ket{+}$ in the second slot, while Bob transmits $\ket{V}$ in the first slot and $\ket{-}$ in the second slot. In Cycle 1, Charlie's detector $D_1$ in the first slot and $D_3$ in the second slot click. Since the two detected states are not in the same slot, there are no coincidence count~(CC). Similar to the case in Cycle 1, $D_2$ in the first slot and $D_4$ in the second slot click in Cycle 2, and there is no CC neither. \par

However, if we consider the detection results of Cycle 1 and Cycle 2 as a merged block, the upper quantum states are from Cycle 1 and the lower states are from Cycle 2 as shown in~\cref{Method for cross-cycle}(b). Detection of $D_1$ in Cycle 1 and detection of $D_2$ in Cycle 2, which are both in the first slot, can be treated as coincidence counts of the Bell-state measurement but in the sampling process. In the second slot, coincidence counts occur owing to $D_3$ and $D_4$. As a result, there are two coincidence counts in this merged block, which are counted into $C_i$. It should be noted that if and only if the two quantum states are in the same slot in one block and they conform to the Bell-state measurement, there will be one cross-cycle count in the $C_i$~\cite{note2}. This method increases the number of coincidences by rebuilding the post-selected Bell state cross cycles, overcoming the limitation of samples. It is noted that we also analyse the possible error propagation due to this cross-cycle method, which luckily shows that there is no impact of error propagation. The details about the error-propagation analysis from two different aspects are presented in the Appendix.\ref{Appendix:A}. \par
\begin{table}[h]
\centering
\renewcommand\arraystretch{1.2}
  \caption{The result of cross-cycle coincidence count, $C_i$, in the MDI system.}
  \begin{center}
  
   \begin{tabular}{c|c c c c}
    \hline
        \diagbox{Pre}{Cur} & $\omega$ & $\nu$ & $\mu$ & $s$ \rule{0pt}{1.2ex}\\
        \hline
        $\omega$ & \num{7.38e5} & \num{3.91e6} & \num{4.97e7} & \num{3.57e7} \\
        $\nu$& \num{4.81e6} & \num{1.58e7} & \num{1.17e7}&\num{1.28e8} \\
        $\mu$& \num{2.90e6} & \num{9.92e6} & \num{7.88e6} & \num{6.03e7} \\
        $s$ & \num{2.45e7}& \num{8.22e7}& \num{4.17e7} & \num{4.03e8}\\
        \hline
    \end{tabular}
  \end{center}
\label{cross-cycle Results}
\end{table}

The outcome of the cross-cycle method is presented in Table.~\ref{cross-cycle Results}. According to the method of cross-cycle coincidence count, the range of detection is increased by the amplification factor of $n_b*n_b$, with $n_b$ cycles in a block. However, in order to reduce the complexity of computation, the calculation program only considers the coincidence counts between the $n$-th and the $(n+i)$-th cycles, which leads that the amplification factor becomes to $n_b*(n_b-1)/2$. Thus, the transmission numbers, $T_i$, also are modified to $T_i$ = $G_i*n_b*(n_b-1)/2$. Since the entire data is too large to be considered as a single block, the \num{1.6e6} cycles are divided into \num{10} blocks, and thereby $n_b=\num{1.6e5}$. In our process, the average number of cross-cycle coincidence among these 10 blocks is used as $C_i$. Consequently, in Stage 4, $R_i =C_i/T_i=C_i/(G_i*n_b*(n_b-1)/2)$.\par

\begin{figure}[h]
\centering
\includegraphics[width=\linewidth]{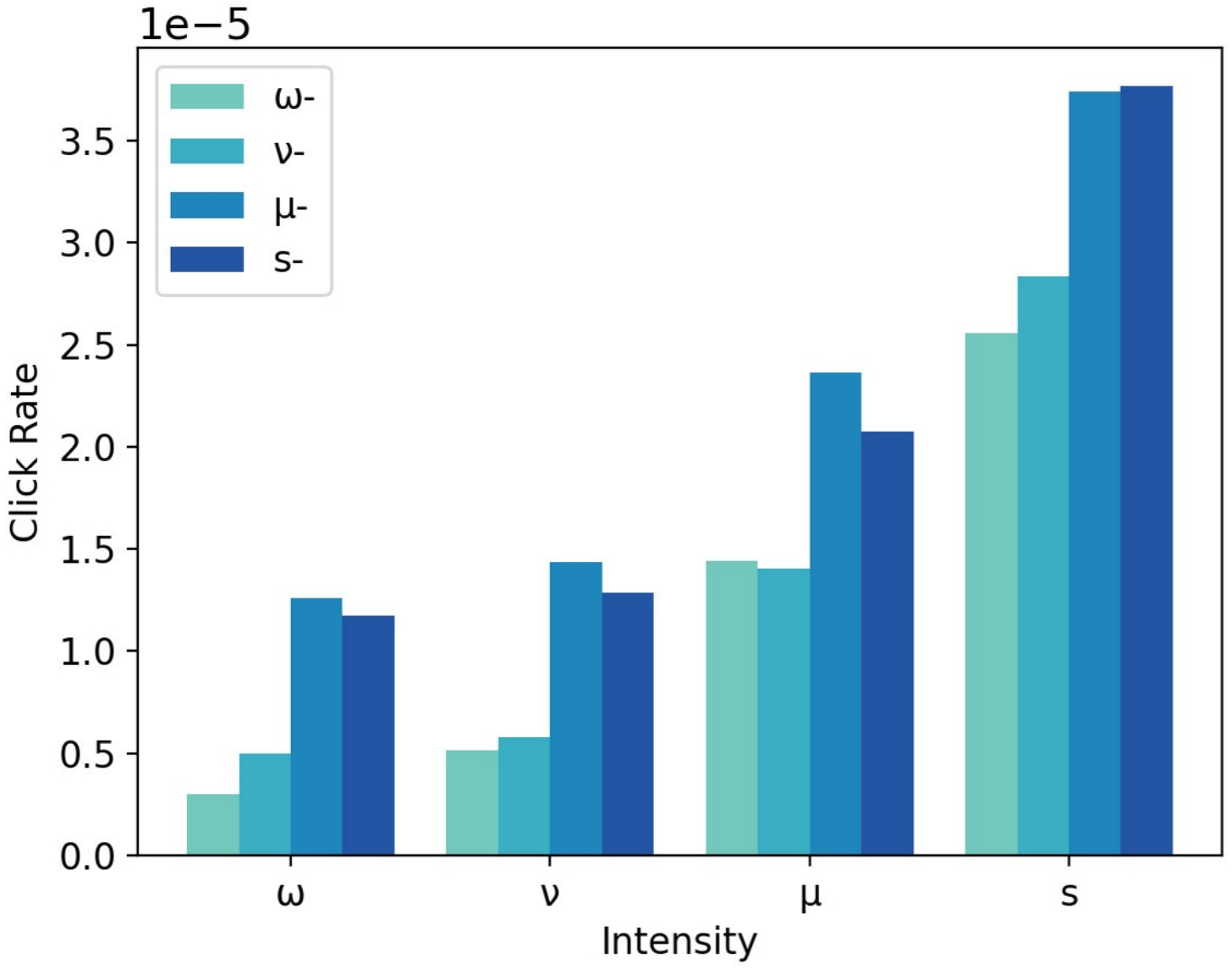}
\captionsetup{font={small,stretch=1.25},justification=raggedright}
\caption{{\bfseries The cross-cycle method for MDI QKD.} \justifying{There are four states for decoy-state protocol, $\omega$ for the vacuum state, $\mu$, $\nu$ for the decoy states and $s$ for the signal state. It shows that $\mu$ and $s$ enhance the intensity of the target pulse at $k$-th while the $\nu$ and $\omega$ reduce the intensity of target pulse.}}
\label{Result for MDI}
\end{figure}
The click rates obtained from the experimental data are displayed in Fig.~\ref{Result for MDI}, where the difference in click rates at the same $k$-th pulse indicates the effect of intensity correlation. For all the intensities of the $k$-th pulse in the MDI QKD system, when the previous intensity is $s$ or $\mu$, the click rate increases; while when the previous intensity is $\nu$ or $\omega$, the click rate decreases. From a general point of view, the largest absolute discrepancy with the same $k$-th pulse is \num{1.2e-5} between $ss$ and $\omega s$, with 37\% relative fluctuation. The vacuum state $\omega$ exhibits the greatest relative fluctuation, about 100\% fluctuation between $\omega\omega$ and $\mu \omega$, whose absolute discrepancy is \num{7.6e-6}. For $\omega$, $\nu$, $\mu$ as the $k$-th pulse, the highest click rates among their own groups is observed as $\mu a$, which is higher than $sa$, $a=\omega, \nu, \mu$. Conversely, $ss$ has the highest click rate for $s$ as the $k$-th pulse. In general, the intensity correlation in the MDI QKD system is pronounced and the detection result can be used to monitor the correlation during the operation of the system.

\section{Effect on the Security of BB84 QKD}
\label{Effect on the Security of QKD}
After employing the measurement method proposed in this work to assess the intensity correlations in QKD systems, we apply these measurement results to the theoretical security analysis of intensity correlations~\cite{zapatero2021security}, allowing us to quantitatively analyze the impact of intensity correlations on the security of the BB84 QKD system. \par 
Firstly, we use the model in Ref.~\cite{huang2022dependency} to convert the click rate into mean photon numbers, which is used to evaluate the secure key rate~(SKR). Specifically, the average click rates for {$V$, $D_1$, $D_2$, $S$} are considered as ${\hat{D}}_{\mu_i}$, with {${\mu}_i$= $V$, $D_1$, $D_2$, $S$} and $P_{{\mu}_i} = 3 : 7 : 35 : 55$. $\tau$ represents the number of gates during each dead-time interval, which is chosen as 5 in our case. Consequently, the overall after-pulse rate $q$= 0.022, according to Ref.~\cite{huang2022dependency}. The mean photon numbers of the quantum states are calculated to be {$\mu_V=0.001$, $\mu_{D_1}=0.03$, $\mu_{D_2}=0.09$, and $\mu_{S}=0.23$}.\par

According to the analysis of Gottesman-Lo-Lütkenhaus-Preskill (GLLP)~\cite{gottesman2004security}, we need to estimate the lower bound of $Y_{1}^{\mu}$ and the upper bound of $e^{\mu}_{1}$ based on the experimentally available parameters. Reference~\cite{zapatero2021security} derives explicit expressions for these two parameters based on a fundamental result from Lo-Preskill's (LP) security analysis~\cite{lo2007security,Pereira2019quantum} and reference technique~\cite{pereira2020quantum}, utilizing the Cauchy-Schwarz~(CS) inequality within the complex Hilbert space. The formula for lower bound of $Y_{1}^{\mu}$ is expressed as~\cite{zapatero2021security}
    \begin{align}
        \text{min} \quad & Y_{1}^{\mu}
        \label{eq:minY}
    \end{align}
\begin{align*}
\text{s.t.} \quad & Q_a \geq e^{-a^+} Y_{0}^{a} + \sum_{n=1}^{n_{\text{cut}}} \frac{e^{-a^-} a^{-n}}{n!}Y_{n}^{a}, \\
& Q_a \leq 1 - e^{-a^+} + e^{-a^-} Y_{0}^{a} - \sum_{n=1}^{n_{\text{cut}}} \frac{e^{-a^+} a^n}{n!} (1 - Y_{n}^{a}), \\
& c_{ab,n}^+ + m_{ab,n}^+ Y_{n}^{a} \geq Y_{n}^{b}, \\
& c_{ab,n}^- + m_{ab,n}^- Y_{n}^{a} \leq Y_{n}^{b}, \\
& 0 \leq Y_{n}^{a} \leq 1,  \\
& (a \in A, b \in A, b \neq a, n = 0, \ldots, n_{\text{cut}})
\end{align*}
where $a,b$ represent any two different intensity settings, while $A$ represents the set of all possible intensity settings. In our experiment, there are four intensity settings: $V,D_1,D_2$ and $S$. Therefore, we have $a,b \in A = \{V, D_1, D_2, S\}$. $Y_n^a$ is yield of $n$-photon pulses when intensity setting is $a$. The expressions for the parameters $c_{ab,n}^+, c_{ab,n}^-, m_{ab,n}^+,$ and $m_{ab,n}^-$ are given using the CS inequality within the complex Hilbert space. Since the actual intensity follows a distribution due to intensity correlation, the parameter $\delta$ represents the relative deviation in the distribution of the state. $\delta_{max}$ is defined as the maximum relative deviation of the distributions among all the intensities settings. Thus, we use the maximum relative deviation among all the intensity settings to describe the fluctuation for intensity settings, which means $a^{\pm} = a (1 \pm \delta_{max})$. Now, we acquire the lower bound of $Y_{1}^{\mu}$, according to the CS inequality, which is also applicable for the upper bound of $e^{\mu}_{1}$~\cite{zapatero2021security}. Additionally, previously experimental study of correlation distribution shows that the actual transmitted intensities is within a range that follows the Gaussian distribution~\cite{li2023high}, which is also found in our experiment. In this case, one can use the truncated Gaussian distribution model to analyze the impact of the measured intensity correlation on the practical security of QKD systems~\cite{sixto2022security}. Specifically, when the intensity distribution under known correlation approximately follows a Gaussian distribution, one can more accurately estimates the number of photons emitted by the transmitter. This allows a tighter bound in the parameter estimation, resulting in a significant improvement in the key rate~\cite{sixto2022security}.

To apply this theoretical model to a quantitative analysis of our experimental results, our primary objective is to identify the value of $\delta_{max}$ from the experimental outcomes. Given that $k=2$ and $p=4$ in our study of BB84 QKD system, there are 16 groups in Stage 1. Since there are 4 corresponding groups for each $a \in \{V,D_1,D_2,S\}$, the 16 groups can be used to obtain the maximum relative deviation of all the $a$, which means that $\delta_{max}$ of $a_2$ can be found from the distribution of $a_1a_2$, with $a_1,a_2 \in \{V,D_1,D_2,S\}$. \par

\begin{figure}[h]
\centering
\includegraphics[width=\linewidth]{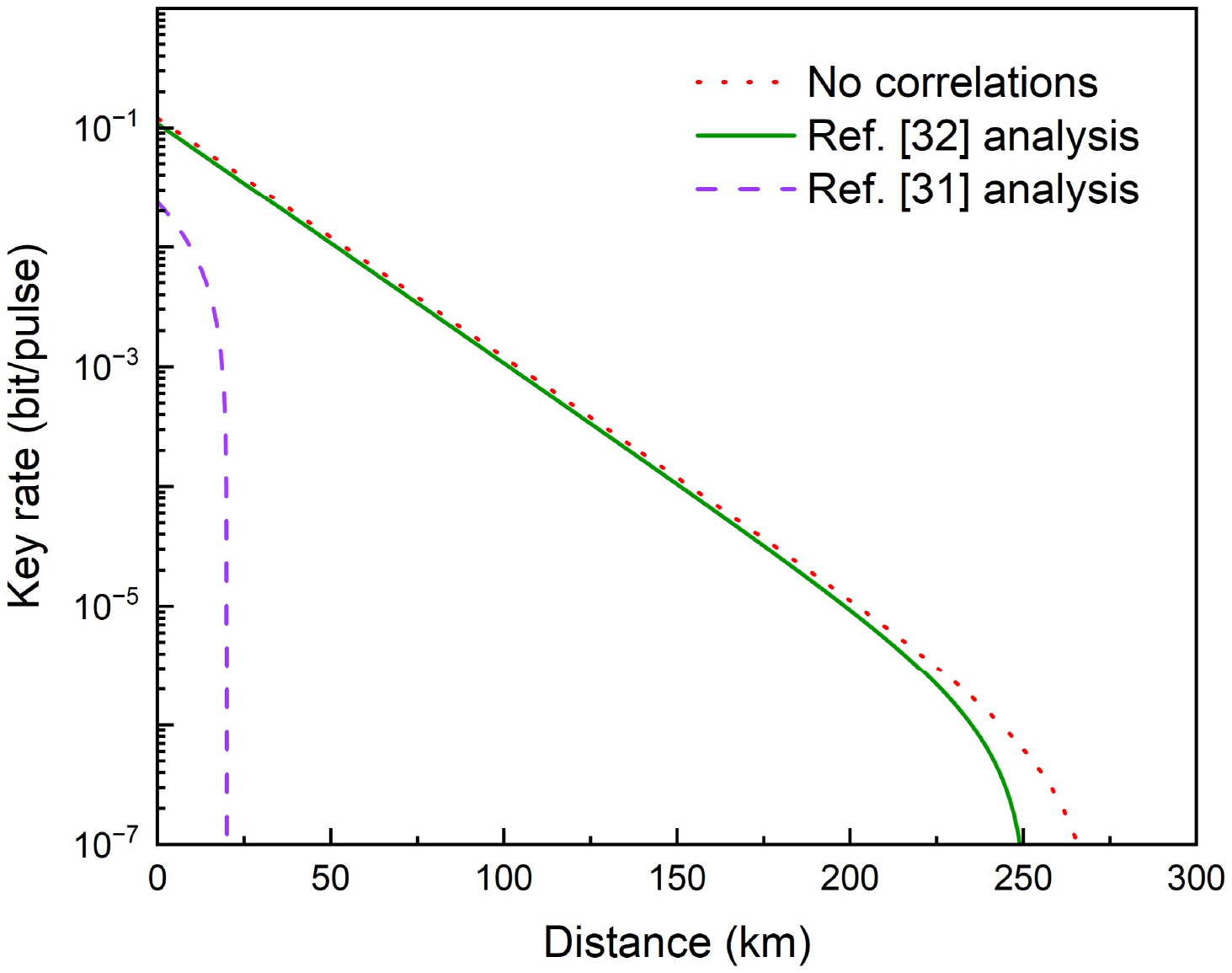}
\caption{ {\bfseries The key generation rate of QKD in the presence of intensity correlation.}\justifying{ The dotted line represents the SKR without intensity correlation, while the dashed and solid lines represent the SKRs with experimental results in two different security frameworks. In the specific simulations, to simplify the calculations, we selected the correlation length $k=2$ and intensity settings $\mu_{S}=0.23$, $\mu_{D_1}=0.03$, and $\mu_V=0.001$.}}
\label{fig:key_rate} 
\end{figure}
We set $\delta= 3 \sigma / \mu$, where $\mu$ represents the mean value of the Gaussian distribution and $\sigma$ is the standard deviation. The intensity fluctuation is affected by two factors -- the correlated previous intensity $a_1$ and the random fluctuations of $a_2$ itself, which are assumed to be independent of each other. Consequently, the fluctuations solely due to the changes in $a_1$ can be calculated by switching the different choice of $a_1$. To illustrate, consider the cases $a_1a_2=SS$ and $a_1a_2=VS$. In $SS$, the actual intensity of $a_2$ follows the Gaussian distribution $N(\mu_{SS},\sigma_{SS}^2)$, whereas in $VS$, it follows $N(\mu_{VS},\sigma_{VS}^2)$. Consequently, the deviation solely due to different $a_1$ can be represented as $\sigma_{VS}^2-\sigma_{SS}^2$, assuming that $\sigma_{VS}^2>\sigma_{SS}^2$. Subsequently, when $a_2=S$ and $a_1$ switches from $S$ to $V$, the range of fluctuations solely due to intensity correlation can be calculated 
    \begin{align}
       a_{VS}^{\pm} = \mu_{VS} \pm 3\sqrt{\sigma_{VS}^2-\sigma_{SS}^2}.
       \label{eq:alpha}
    \end{align}    

According to Eq.~\ref{eq:alpha}, we calculate the fluctuation of all the groups, using the experimental data from the BB84 QKD system. This reveals that the maximum relative deviation $\delta_{max} = 0.63$ under the consideration that $k=2$ for the length of correlation. Subsequently, $\delta_{max}$ and $k$ are substituted into the security model described above to calculate the SKR, which is shown as in Fig.~\ref{fig:key_rate}. The experimental tests and security theoretical analysis demonstrate that intensity correlation reduces the secure key rate of the QKD systems.

\section{Conclusion}
\label{Conclusion}
In this study, we propose a method to measure the intensity correlation in the QKD system via the single-photon detectors equipped in the measurement apparatus. Unlike the previous research that employs extra classical optical detectors, our method  directly utilizes the detection results in the measurement unit, which introduces a more practical solution to characterize the intensity correlations in a QKD system without modifying its configuration. This methodology is applied to the BB84 QKD system and the MDI QKD system respectively, evaluating the intensity correlation of the quantum states in these systems. The experimental result demonstrates that both QKD systems exhibit significant effect of intensity correlation. Moreover, according to the security analysis, the intensity correlation decreases the secret key rate, which may provide Eve an approach to eavesdrop the information.\par

It is shown that the methodology proposed in this work is applicable to diverse QKD protocols, since it accounts for multiple factors, including the number of source units (denoted as $l$), distinct intensity correlation lengths (denoted as $k$), and various intensity choices (denoted as $p$). Although we only show the examples to analyze the correlation of the nearest neighbouring quantum states, the method can be further employed in higher-order correlation analysis for $k > 2$. Most importantly, our method enables the simultaneous monitoring of intensity correlations in parallel with the operation of a QKD system, because the data required by this method can be obtained during the raw key exchange. It is notable that this methodology of characterizing the intensity correlation of a QKD system can be applied to the standard security certification of a QKD system. Whereas, the security framework considering intensity correlation in the MDI QKD scheme still shall be fully developed to quantify the effect of intensity correlation in the MDI QKD implementation. \par

\section*{Funding.}National Natural Science Foundation of China (Grant No. 62371459, 12174374, and 12274398); Innovation Program for Quantum Science and Technology (2021ZD0300704); National Key Research and Development Program of China (2019QY0702).

\section*{Acknowledgement.} We thank Marcos Curty, and Hongsong Shi, and Fengyu Lu for helpful discussions.

\section*{Disclosures.}The authors declare no conflicts of interest.
\section*{Data availability.} Data underlying the results presented in this paper are not publicly available at this time but maybe obtained from the authors upon reasonable request.

\appendix

\section*{Appendix}
\section{Uncertainty analysis of cross-cycle method}
\label{Appendix:A}
\begin{figure}[h]
\centering
\includegraphics[width=\linewidth]{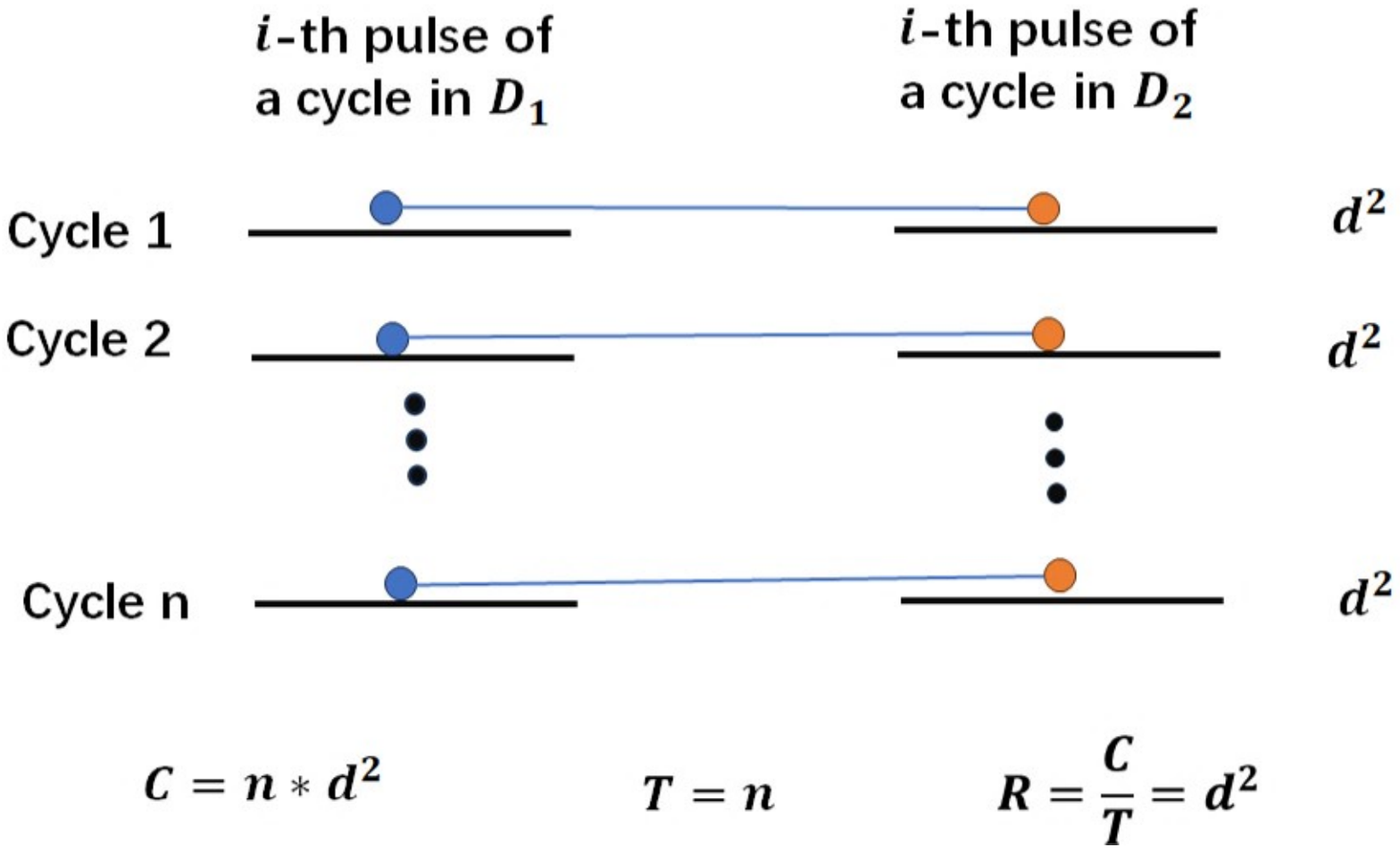}
\captionsetup{font={small,stretch=1.25},justification=raggedright}
\caption{{\bfseries The original coincidence counts.} \justifying{The coincidence counts of two detectors $D_1$ and $D_2$ when the clicks from $D_1$ and $D_2$ are counted in the same cycle.}}
\label{fig:Orignal CC}
\end{figure}

First, we calculate the original click rate of the coincidence counts. Assuming the click rate for a single-photon detector is $d$, and the coincidence counts are caused by the clicks from the two individual detectors after sifting. Without prejudice of the generality, we only consider the $i$-th pulse in each detection cycle as Fig.~\ref{fig:Orignal CC} shown. For the coincidence counts from $D_1$ and $D_2$ at the $i$-th slot in a MDI QKD system, the click rate is $d^2$. Thus, for $n_b$ cycles, $C=n_b* d^2$ and $T=n_b$ in total at the same $i$-th slot. Consequently, the average click rate for the original coincidence counts is $R = d^2$. The error (or say the uncertainty) of the click rate is assumed to be $e$.

Regarding the cross-cycle method, we then analyse the error in two ways, cycle-by-cycle calculation and click-numbers-based categorization.

\begin{figure}[h]
\centering
\includegraphics[width=\linewidth]{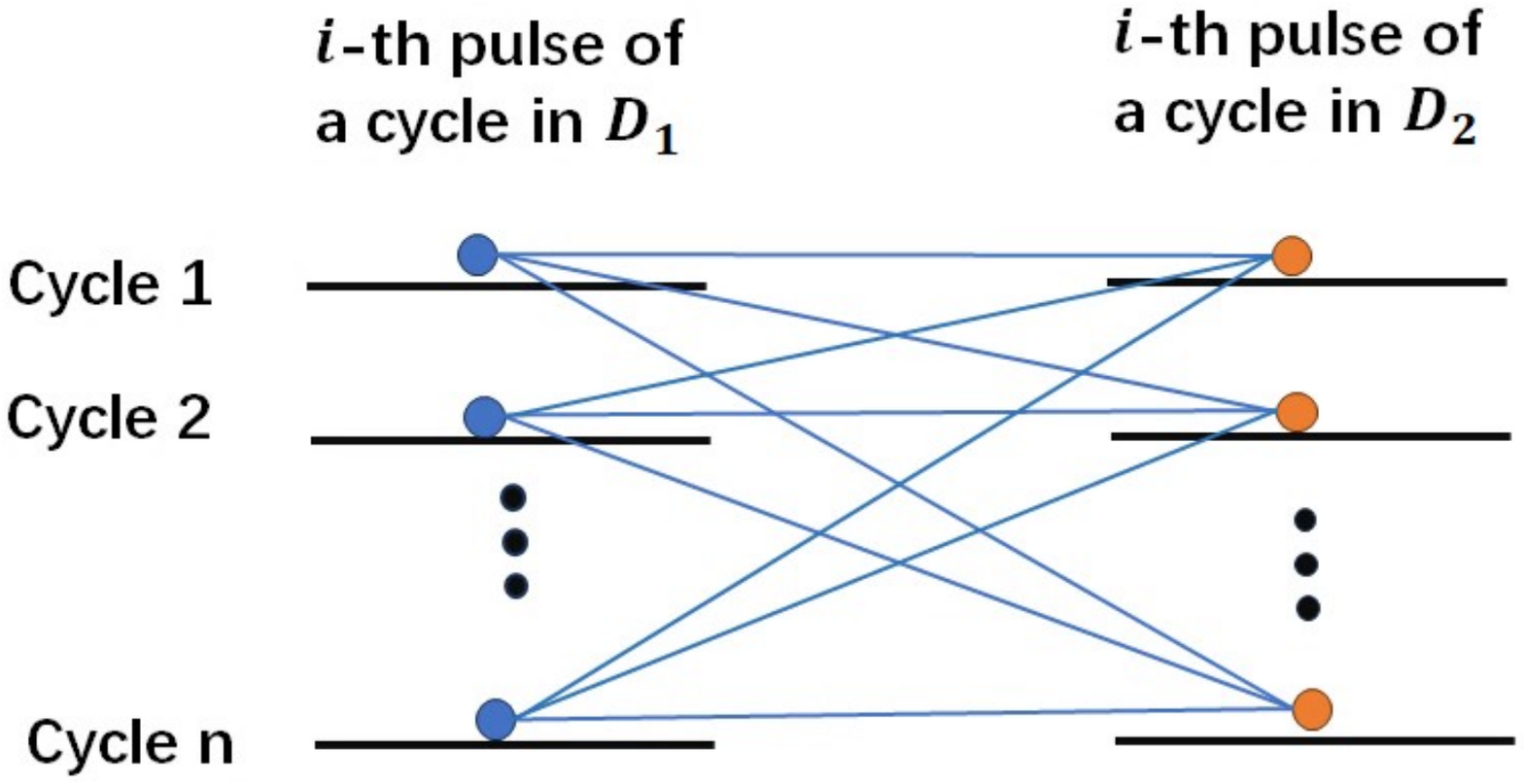}
\captionsetup{font={small,stretch=1.25},justification=raggedright}
\caption{{\bfseries The cross-cycle method} \justifying{The click in $D_1$ can match all the clicks in $D_2$.}}
\label{fig:CC-2}
\end{figure}

{\bf (a)~Cycle-by-cycle calculation.} We also assume that there are $n_b$ cycles in each block and consider the click result of the $i$-th pulses. The cross-cycle method indicates that each click in $D_1$ can match all the clicks in $D_2$, shown as in Fig.~\ref{fig:CC-2}. For each cycle, the click rate is $d$. Due to the cross-cycle method, for each click in $D_1$, the expectation of its matches in $D_2$ is $n_b*d$, which means there are $n_b*d^2$ matches for each cycle in $D_1$. Since there are $n_b$ cycles for $D_1$, the expectation of the matches in a block is $n_b*n_b*d^2$= $n_b^2*d^2$. As for $T$, when every pair of pulses causes coincidence clicks, the result is $n_b*n_b$. Thus, $R=C/T =d^2$, the same as the original coincidence counts. From this aspect of view, the error (or say the uncertainty) for cross-cycle method is the same as that of the original coincidence count, which is also $e$. \par

{\bf (b)~Click-number-based categorization.} Apart from cycle-by-cycle calculation, we also cross check the correctness based on the click-number categorization. The main idea of this method can be summarized as follows. First, categorize the different cases based on the total number of clicks from Alice and Bob. Then, within each category, we discuss the possible numbers of coincidence counts and their corresponding probabilities. Finally, we calculate the expected value of click rate according to all the possible numbers of coincidence counts based on their probabilities.

Categorizing Alice and Bob by their total number of the clicks among $n_b$ cycles, there are a total of $2n_b+1$ cases. That is, Alice and Bob could have different cases of 0, 1, ..., $2n_b-1$, $2n_b$ clicks in total. A discussion of the sub-cases is as follows.

(1). For the case that there is 0 click among Alice and Bob, i.e., Alice has 0 click and Bob has 0 click, there is 0 coincidence count with probability of $C_{n_b}^0d^0(1-d)^{n_b}*C_{n_b}^0d^0(1-d)^{n_b}$.

(2). For the case that there is 1 click among Alice and Bob,
\par
    \hspace*{0.5cm}a. if Alice has 0 click and Bob has 1 click, there is 0 coincidence count with probability of $C_{n_b}^0d^0(1-d)^{n_b}*C_{n_b}^1d^1(1-d)^{n_b-1}$.

    \hspace*{0.5cm}b. if Alice has 1 click and Bob has 0 click, there is 0 coincidence count with probability of $C_{n_b}^1d^1(1-d)^{n_b-1}*C_{n_b}^0d^0(1-d)^{n_b}$.
    
(3). For the case that there are 2 clicks among Alice and Bob,
\par
    \hspace*{0.5cm}a. if Alice has 0 click and Bob has 2 clicks, there is 0 coincidence count with probability of $C_{n_b}^0d^0(1-d)^{n_b}*C_{n_b}^2d^2(1-d)^{n_b-2}$.

    \hspace*{0.5cm}b. if Alice has 1 click and Bob has 1 click, there is 1 coincidence count with probability of $C_{n_b}^1d^1(1-d)^{n_b-1}*C_{n_b}^1d^1(1-d)^{n_b-1}$.

   \hspace*{0.5cm}c. if Alice has 2 clicks and Bob has 0 click, there is 0 coincidence count with probability of $C_{n_b}^2d^2(1-d)^{n_b-2}*C_{n_b}^0d^0(1-d)^{n_b}$.

(4). For the case that there are 3 clicks among Alice and Bob, 
\par
    \hspace*{0.5cm}a. if Alice has 0 click and Bob has 3 clicks, there is 0 coincidence count with probability of $C_{n_b}^0d^0(1-d)^{n_b}*C_{n_b}^3d^3(1-d)^{n_b-3}$.

    \hspace*{0.5cm}b. if Alice has 1 click and Bob has 2 clicks, there are 2 coincidence counts with probability of $C_{n_b}^1d^1(1-d)^{n_b-1}*C_{n_b}^2d^2(1-d)^{n_b-2}$.

    \hspace*{0.5cm}c. if Alice has 2 clicks and Bob has 1 click, there are 2 coincidence counts with probability of $C_{n_b}^2d^2(1-d)^{n_b-2}*C_{n_b}^1d^1(1-d)^{n_b-1}$.

    \hspace*{0.5cm}d. if Alice has 3 click and Bob has 0 click, there is 0 coincidence count with probability of $C_{n_b}^3d^3(1-d)^{n_b-3}*C_{n_b}^0d^0(1-d)^{n_b}$.

By following this regulation, the number of coincidence counts and the corresponding probability in each case can be calculated, and the last case is as follows.\par
(2$n_b$+1). For the case that there are $2n_b$ clicks among Alice and Bob, i.e., Alice has $n_b$ clicks and Bob has $n_b$ clicks, there are $n_b^2$ coincidence counts with probability of $C_{n_b}^{n_b}d^{n_b}(1-d)^{0}*C_{n_b}^{n_b}d^{n_b}(1-d)^{0}$.

Summing up all the cases of coincidence counts with the condition of probability, which is calculated by a simulation program, we obtain the result numerically equals to $n_b^2*d^2$. Additionally, it is obvious that $T=n_b^2$, when all the $i$-th pulses of Alice and Bob in $n_b$ cycles click. Thus, $R= C/T=d^2$. The error (or say the uncertainty) is also the same as that of the original coincidence count.
It is important to note that, in order to simplify the analysis process, the aforementioned two methods consider the entire cross-cycle process with the amplification factor of $n_b*n_b$. Although in the application case of MDI QKD, the practical process only considers roughly half of cross-cycle coincidence counts, the outcome and the error for $R_i$ are consistent with the above analysis.

According to the above analysis from two different points of view, there is no impact of error propagation in the proposed cross-cycle method.


\begin{thebibliography}{43}
\bibitem [{not({\natexlab{a}})}]{note1}%
  It is noted that the aforementioned counting method does not account for the correlation between the final intensity in the previous string and the first intensity in the following string. Since this case happens in one of 10,000 groups, it has a minimal impact on the results.%
\bibitem [{not({\natexlab{b}})}]{note2}%
  For each intensity setting of the target source, all possible coincidence counts consistent with the Bell-state measurement are collected in $C_i$, indicating that the effect from the non-target source plays the same role in each intensity setting of the target source. Consequently, the results of the Bell state measurement can be used to represent the intensity correlation of the target source.%
\end{thebibliography}
%

\end{document}